\documentclass{article}
\begin{document}
\hoffset = -2 truecm \voffset = -2 truecm
\newcommand{\dfrac}[2]{\frac{\displaystyle #1}{\displaystyle #2}}
\title{Real-Time thermal Ward-Takahashi Identity for vectorial current in QED and QCD
\footnote{This work was
supported by the National Natural Science Foundation of China.} \\
}
\author{Bang-Rong Zhou\\
 Department of Physics, Graduate School of the Chinese Academy of Sciences,\\
Beijing 100039, China  and \\
 CCAST (World Laboratory) P.O. Box 8730, Beijing 100080, China}
\date{(November 12, 2003)}
\maketitle
\begin{abstract}
It is shown that , by means of canonical operator approach, the Ward-Takahashi
identity (WTI) at finite temperature $T$ and finite chemical potential $\mu$ for
complete vectorial vertex and complete fermion propagator can be simply proven,
rigorously for Quantum Electrodynamics (QED) and approximately for Quantum
Chromodynamics (QCD) where the ghost effect in the fermion sector is neglected. The
WTI shown in the real-time thermal matrix form will give definite thermal
constraints on the imaginary part of inverse complete Feynman propagator including
self-energy for fermion and will play important role in relevant physical processes.
When the above inverse propagator is assumed to be real, the thermal WTI will
essentially be reduced to its form at $T=\mu=0$ thus one can use it in the latter's
form. At this point, a practical example is indicated.
\end{abstract}
\indent Ward-Takahashi identity(WTI) \cite{kn:1,kn:2} is a relation of fundamental
importance in field theory, including in a thermal gauge theory. For instance, in
thermal QCD, when discussing the Schwinger-Dyson equation for guark self-energy and
the effective potential for quark propagator, one needs a non-perturbative WTI for
complete vectorial vertex and complete fermion propagator, since in these cases the
used fermion propagators are complete ones without perturbative corrections. If we
use the real-time  formalism of thermal field theory, then the WTI should be
expressed in a thermal matrix form. On the WTI in the real-time thermal field
theory, many researches have been made, e.g. a general demonstration of WTI from
canonical operator approach of fields \cite{kn:3}, a formal discussion of WTI from
the path integration approach \cite{kn:4} and an explicit check of WTI from a
perturbation theory \cite{kn:5} etc.. However, these researches did not involve the
explicit matrix expression of the above non-perturbative WTI. Hence, for practical
purpose, it is interesting to make a deep-going investigation of the
non-perturbative WTI. The rigorous thermal WTI in a gauge theory, in principle, may
be derived from Becchi-Rouet-Stora (BRS) invariance \cite{kn:6}. However, when the
problem is confined to the WTI for vectorial vertex in a vectorial gauge theory, we
can alternatively show a more simple approach to derive it. In this paper, we will
indicate that, following the same approach to obtain the WTI in zero temperature
QED, the non-perturbative WTI at a finite temperature $T$ and a finite chemical
potential $\mu$ can be proven simply by means of the operator formalism of the
fields, rigorously for QED and approximately for QCD. Then the feature of the
resulting WTI and its relation to practical application will be analyzed further. \\
\indent First let us consider the case of temperature QED. In QED, gauge invariance
amounts to electromanetic current conservation, hence the WTI at finite $T$ and
$\mu$ can also be derived from the latter. In the real-time formalism of thermal
field theory\cite{kn:7}, the conserved Noether vectorial currents $j^{(r)}_{\nu}(x)$
may be expressed by
\begin{equation}
 j^{(r)}_{\nu}(x)=\bar{\psi}^{(r)}(x)\gamma_{\nu}\psi^{(r)}(x), \; r=1,2
\end{equation}%%(1)%%
which obeys
\begin{equation}
\partial^{\nu}j^{(r)}_{\nu}(x)=0, \; r=1,2,
\end{equation}%%(2)%%
where $\psi^{(r)}(x)$ are the fermion field operators, $r=1$ and $2$ respectively
denote physical and thermal ghost fields (here the attribute "thermal" has been
added so as to distinguish the ghost in thermal field theory from the ghost in a
gauge theory).  Now define
\begin{equation}
j_{\nu}(x)=\sum_{r=1,2}\bar{\psi}^{(r)}(x)\gamma_{\nu}\psi^{(r)}(x),
\end{equation}%%(3)%%
then it is also conserved, i.e.
\begin{equation}
\partial^{\nu}j_{\nu}(x)=0.
\end{equation}%%(4)%%
We first prove that conservation of the current $j_{\nu}(x)$ will lead to the
equality
\begin{equation}
\partial^{\nu}_x\texttt{T}j_{\nu}(x)\psi^{(r)}(x_1)\bar{\psi}^{(s)}(x_2)=
[-\delta^4(x-x_1)+\delta^4(x-x_2)]\texttt{T}\psi^{(r)}(x_1)\bar{\psi}^{(s)}(x_2),
 r, s =1, 2,
\end{equation}%%(5)%%
where $\partial^{\nu}_x$ represents only the partial derivative for $x$ and
$\texttt{T}$ is the time-order operation. In fact, by means of the definition of
 the $\texttt{T}$ product  and Eq.(4), we can obtain
\begin{eqnarray*}
  \partial^{\nu}_x\texttt{T}j_{\nu}(x)\psi^{(r)}(x_1)\bar{\psi}^{(s)}(x_2) &=&
  \delta(x^0-x_1^0)\texttt{T}[j_0(x),\psi^{(r)}(x_1)]\bar{\psi}^{(s)}(x_2) \\
  &&
  +\delta(x^0-x_2^0)\texttt{T}\psi^{(r)}(x_1)[j_0(x),\bar{\psi}^{(s)}(x_2)].
\end{eqnarray*}
Considering the fact that the physical fields and the thermal ghost fields commute
(or anticommute) each other \cite{kn:7}, we will always have
\begin{eqnarray*}
&&\left.[j_0(x),\psi^{(r)}(x_1)]\right|_{x^0=x_1^0}=-\delta^3(\vec{x}-\vec{x_1})\psi^{(r)}(x_1), \\
&&\left.[j_0(x),\bar{\psi}^{(s)}(x_2)]\right|_{x^0=x^0_2}=\delta^3(\vec{x}-\vec{x_2})\bar{\psi}^{(s)}(x_2),
\end{eqnarray*}
and then are led to Eq. (5) being proven.
\\
\indent Now taking thermal expectation values of the both sides of Eq.(5) and
defining the thermal three-point function
\begin{equation}
G^{rs}_{\nu}(x;x_1,x_2)=\langle
0|\texttt{T}j_{\nu}(x)\psi^{(r)}(x_1)\bar{\psi}^{(s)}(x_2)|0\rangle_T
\end{equation}%%(6)%%
and the thermal propagator
\begin{equation}
G^{rs}(x_1,x_2)=\langle 0|\texttt{T}\psi^{(r)}(x_1)\bar{\psi}^{(s)}(x_2)|0\rangle_T
\end{equation}%%(7)%%
where $\langle0|\cdots |0\rangle_T$ represents thermal expectation value, then we
will get
\begin{equation}
\partial^{\nu}_xG^{rs}_{\nu}(x;x_1,x_2)=
[-\delta^4(x-x_1)+\delta^4(x-x_2)]G^{rs}(x_1-x_2), \;
 r, s =1, 2.
\end{equation}%%(8)%%
In Eq. (8) the translation invariance of $G^{rs}(x_1,x_2)$ has been considered. The
Fourier transformation of $G^{rs}_{\nu}(x;x_1,x_2)$ and $G^{rs}(x_1-x_2)$ are
denoted  respectively by $\tilde{G}^{rs}_{\nu}(q;p_1,p_2)$ and $iS'^{rs}(p_1)$ or
$iS'^{rs}(p_2)$, where $\tilde{G}^{rs}_{\nu}(q;p_1,p_2)$ is the complete vectorial
vertex and $iS'^{rs}(p)$ is the complete fermion propagator.  Then the Fourier
transformation of Eq.(8) will lead to that
\begin{equation}
iq^{\nu}\tilde{G}^{rs}_{\nu}(q;p_1,p_2)=iS'^{rs}(p_2)-iS'^{rs}(p_1),\;\; q=p_2-p_1.
\end{equation}%%(9)%%
On the other hand, $G^{rs}_{\nu}(x;x_1,x_2)$ is the three-point function with
external propagators (legs),  so it can be written by
\begin{equation}
G^{rs}_{\nu}(x;x_1,x_2)=iS'^{rt}(x_1-x)\Gamma_{\nu}^{tt'}(x)iS'^{t's}(x-x_2).
\end{equation}%%(10)%%
The Fourier transformation of Eq.(10) leads to that
\begin{equation}
\tilde{G}^{rs}_{\nu}(q;p_1,p_2)=iS'^{rt}(p_1)\Gamma_{\nu}^{tt'}(p_2,p_1)iS'^{t's}(p_2),
\; \; q=p_2-p_1.
\end{equation}%%(11)%%
Substituting Eq.(11) into Eq.(9) we obtain
\begin{equation}
iS'^{rt}(p_1)iq^{\nu}\Gamma_{\nu}^{tt'}(p_2,p_1)iS'^{t's}(p_2)=iS'^{rs}(p_2)-iS'^{rs}(p_1).
\end{equation}%%(12)%%
In the form of thermal matrix, Eq. (12) can be expressed by
\begin{equation}
i\hat{S}'(p_1)iq^{\nu}\hat{\Gamma}_{\nu}(p_2,p_1)i\hat{S}'(p_2)=i\hat{S}'(p_2)-i\hat{S}'(p_1).
\end{equation}%%(13)%%
The thermal matrix propagator $i\hat{S}'(p)$ is defined by \cite{kn:7}
$$
i\hat{S}'(p)=\hat{M}_p\left(\matrix{iS_F(p)& 0 \cr
                                           0      &-iS_F^*(p)\cr}\right)\hat{M}_p,
$$
where
$$ \hat{M}_p=\left(\matrix{\cos \theta_p & -e^{\beta \mu/2}\sin \theta_p\cr
                        e^{-\beta\mu/2}\sin \theta_p & \cos \theta_p\cr}\right)
$$
is the thermal transformation matrix with fermionic chemical potential $\mu$,
$\beta=1/T$ ,
$$
\sin^2\theta_p=\frac{\theta(p^0)}{e^{\beta(p^0-\mu)}+1}+\frac{\theta(-p^0)}{e^{\beta(-p^0+\mu)}+1}
$$
and $iS_F(p)$ and $-iS^*_F(p)$ are respectively the complete Feynman propagator for
fermion and its complex conjugate, but the complex conjugate $"^*"$ does not include
the $\gamma$ matrix, in the light of the convention in the real-time thermal field
theory \cite{kn:8}. The thermal vertex matrix $\hat{\Gamma}_{\nu}(p_2,p_1)$ is
defined by
$$ \hat{\Gamma}_{\nu}(p_2,p_1)=\left(\matrix{
                                     \Gamma_{\nu}^{11}(p_2,p_1)&\Gamma_{\nu}^{12}(p_2,p_1) \cr
                                     \Gamma_{\nu}^{21}(p_2,p_1)
                                     &\Gamma^{22}_{\nu}(p_2,p_1)\cr}\right).
$$
The matrix equation (13) can be transformed to
\begin{equation}
q^{\nu}\hat{\Gamma}_{\nu}(p_2,p_1)=[\hat{S}'(p_2)]^{-1}-[\hat{S}'(p_1)]^{-1}.
\end{equation}%%(14)%%
Eq.(13) or Eq.(14) is the WTI in thermal matrix form at finite $T$ and $\mu$ in QED.
They are the rigorous expressions of the thermal WTI for complete vectorial vertex
and complete fermion propagator and obviously, should also  be valid to each order
of the perturbation theory. \\
\indent The WTI at finite $T$ and $\mu$ expressed by Eq.(13) or Eq.(14) which was
derived from the Abelian $U(1)$ current conservation in QED can not be simply
generalized to the non-Abelian $SU_c(3)$ case in QCD. In QCD the corresponding
identity should be the Slavnov-Taylor identity (STI) coming from invariance under
BRS transformation. However, as seen from the case of STI at $T=\mu=0$, if one
neglects the effect of ghosts in the fermion sector \cite{kn:9, kn:10}, and this
amounts to replacing the composite fermion-ghost vertex and the ghost propagator in
the STI by their bare or the lowest order values, the STI for fermion-fermion-gluon
vertex will be reduced to the form of the WTI in QED, only a $SU_c(3)$ generator
matrix must be multiplied in the both sided of the identity \cite{kn:11}. The main
purpose of this approximation is for simplifying practical calculations e.g. making
the derived Schwinger-Dyson equation of fermion self-energy become tractable
\cite{kn:9,kn:10,kn:12}. We will prove that the approximative form of the STI can be
obtained simply by the conservation of the $SU_c(3)$ vector current, hence
neglecting the ghost effect in fermion sector only implies that the local QCD gauge
invariance now becomes the requirement that the color current is conserved at
fermion-fermion-gluon vertex. In fact, when the ghost fields are neglected, the
Lagrangian of QCD including the gauge fixing term is invariant under a global
$SU_c(3)$ transformation and this will lead to conservation of the color current
\begin{equation}
\partial^{\nu}J^a_{\nu}(x)= 0, \;\;
J^a_{\nu}(x)=\bar{\psi}(x)\lambda^a\gamma_{\nu}\psi(x)
    +f^{abc}A^{b\mu}(x)F^c_{\mu\nu}(x),\;  a,b,c=1, \cdots, 8.
\end{equation}%%(15)%%
where  $\lambda^a (a=1, \cdots, 8)$ are the $SU_c(3)$ generators, $f^{abc}$ are the
structure constants of $SU_c(3)$ group, and $A^{b\mu}$ and $F^c_{\mu\nu}$ are the
gauge fields and their field strengths. Eq.(15) was derived in the approximation
that the ghost fields are completely neglected.  However, we will use it only in the
case of fermion-fermion-vertex, i.e. the ghost effect is neglected only in fermion
sector. Thus by means of Eq.(15) and the same demonstrations as that made in QED, we
may obtain the WTI in QCD with the similar form to the one in QED.
\\
\indent Assuming the above approximation is kept at finite $T$ and $\mu$, then one
will be able to derive the thermal WTI for fermion-fermion-gluon vertex in the same
way as the above in QED. As a non-Abelian gauge theory, QCD is normally described by
path integral approach; however, when considered as a field system with constraints,
it may also be discussed in principle by canonical operator approach. In particular,
for derivation of the WTI for fermion-fermion-gluon vertex, one mainly uses the
communicators between the fermion fields and does not involve any communicators
between the gauge fields.
Hence the operator approach is completely feasible in this problem. \\
\indent To derive the thermal WTI for fermion-fermion-gluon vertex in the
approximation of neglecting the ghost effect, we may replace the Abelian vector
current (3) by the non-Abelian $SU_c(3)$ vectorial current
\begin{equation}
j^a_{\nu}(x)=\sum_{r=1,2}\left[\bar{\psi}^{(r)}(x)\lambda^a\gamma_{\nu}\psi^{(r)}(x)
+f^{abc}A^{b\mu(r)}(x)F^{c(r)}_{\mu\nu}(x)\right] , a=1, \cdots, 8,
\end{equation}%%(16)%%
where for simplicity,  we will consider only a single flavor of fermions. By means
of the current conservation $\partial^{\nu}j^a_{\nu}(x)=0$ and the similar
derivation leading to Eq.(5) we can prove that
\begin{eqnarray}
\partial^{\nu}_x\texttt{T}j^a_{\nu}(x)\psi_i^{(r)}(x_1)\bar{\psi}^{(s)j}(x_2)&=&
-\delta^4(x-x_1)(\lambda^a)_i^{\ l}\texttt{T}\psi_l^{(r)}(x_1)\bar{\psi}^{(s)j}(x_2)
\nonumber\\
&&+\delta^4(x-x_2)\texttt{T}\psi_i^{(r)}(x_1)\bar{\psi}^{(s)k}(x_2)(\lambda^a)_k^{\
j}, \;\;\; r, s =1, 2,
\end{eqnarray}%%(17)%%
where $i$ and $j$ denote the color components of the spinors $\psi^{(r)}$ and
$\bar{\psi}^{(s)}$ and we have considered the fact that
$$[f^{abc}A^{bi(r')}(x)F^{c(r')}_{i0}(x), \psi^{(r)}(x_1)]=
[f^{abc}A^{bi(r')}(x)F^{c(r')}_{i0}(x), \bar{\psi}^{(s)}(x_2)]=0.
$$
Noting that the thermal propagators
$$\langle
0|\texttt{T}\psi_l^{(r)}(x_1)\bar{\psi}^{(s)j}(x_2)|0\rangle_T=0, \ \ {\rm when} \ \
l\neq j, $$
$$\langle
0|\texttt{T}\psi_i^{(r)}(x_1)\bar{\psi}^{(s)k}(x_2)|0\rangle_T=0, \ \ {\rm when} \ \
i\neq k,
$$
though they may be non-zeros when $r\neq s$ and $l=j$ or $i=k$. Hence the thermal
expectation value of Eq.(17) will become
\begin{equation}
\partial^{\nu}_x
\langle 0|\texttt{T}j^a_{\nu}(x)\psi_i^{(r)}(x_1)\bar{\psi}^{(s)j}(x_2)|0\rangle_T=
(\lambda^a)_i^{\ j}[-\delta^4(x-x_1)+\delta^4(x-x_2)]
\langle0|\texttt{T}\psi^{(r)}(x_1)\bar{\psi}^{(s)}(x_2)|0\rangle_T, \;
 r, s =1, 2.
\end{equation}%%(18)%%
Since the $SU_c(3)$ generator matrix $\lambda^a$ has been factorized out in the
right-handed side of Eq. (18), we can define
\begin{equation}
\langle0|\texttt{T}j^a_{\nu}(x)\psi_i^{(r)}(x_1)\bar{\psi}^{(s)j}(x_2)|0\rangle_T=
(\lambda^a)_i^{\ j}G^{rs}_{\nu}(x;x_1,x_2).
\end{equation}%%(19)%%
We indicate that since the gauge field sector of $j^a_{\nu}(x)$ has no contribution
to the discussed three-point Green function, the $G^{rs}_{\nu}(x;x_1,x_2)$ defined
by Eq.(19) is actually the fermion-fermion-gluon vertex. As a result of
Eqs.(18)-(19) and Eq.(7), we will have
\begin{equation}
\partial^{\nu}_x(\lambda^a)_i^{\ j}G^{rs}_{\nu}(x;x_1,x_2)=(\lambda^a)_i^{\ j}
[-\delta^4(x-x_1)+\delta^4(x-x_2)]G^{rs}(x_1-x_2),\;\;\;
 r, s =1, 2.
\end{equation}%%(20)%%
Eliminating $\lambda^a$ from the both sides of Eq.(20) we will come to Eq.(8) once
again. This implies that, for the fermion of single flavor and single color, when
the ghost effect is neglected, the WTI for fermion-fermion-gluon vertex in QCD is
identical to that in QED, i.e. the thermal WTI (13) or (14) are also valid in QCD in
the approximation that the ghost effect in fermion sector is neglected. It is
emphasized that the thermal WTI (13) or (14) rigorously for QED and approximately
for QCD are proven by the operator approach here and it seems that the similar proof
did not appear in the literature before. \\
 By means of the explicit matrix expression of Eq.(14)
\begin{eqnarray}
&&\left(\matrix{q^{\nu}\Gamma_{\nu}^{11}(p_2,p_1)&q^{\mu}\Gamma_{\nu}^{12}(p_2,p_1)
\vspace{0.3cm}\cr
 q^{\nu}\Gamma_{\nu}^{21}(p_2,p_1)&q^{\nu}\Gamma^{22}_{\nu}(p_2,p_1)\vspace{0.3cm}\cr}\right)=\nonumber \\
&&\left(\matrix{S_F^{-1}(p)-\sin^2\theta_p[S_F^{-1}(p)-S_F^{*-1}(p)],&
e^{\beta\mu/2}\cos \theta_p\sin \theta_p[S_F^{-1}(p)-S_F^{*-1}(p)]\vspace{0.3cm}\cr
-e^{-\beta\mu/2}\cos \theta_p\sin \theta_p[S_F^{-1}(p)-S_F^{*-1}(p)]&
-S_F^{*-1}(p)-\sin^2\theta_p[S_F^{-1}(p)-S_F^{*-1}(p)]\cr}\right)_{p=p_2}
-(p_2\rightarrow p_1),\nonumber \\
\end{eqnarray}%%(21)%%
we may obtain respectively from its (11), (22) and (12) or (21) components the
equations
\begin{equation}
q^{\nu}\Gamma_{\nu}^{11}(p_2,p_1)=S_F^{-1}(p_2)-S_F^{-1}(p_1)
-\sin^2\theta_{p_2}[S_F^{-1}(p_2)-S_F^{*-1}(p_2)]
+\sin^2\theta_{p_1}[S_F^{-1}(p_1)-S_F^{*-1}(p_1)],
\end{equation}%%(22)%%
\begin{equation}
\Gamma_{\nu}^{22}(p_2,p_1)=-[\Gamma_{\nu}^{11}(p_2,p_1)]^*
\end{equation}%%(23)%%
and
\begin{equation}
q^{\nu}\Gamma_{\nu}^{12}(p_2,p_1)=-e^{\beta\mu}q^{\nu}\Gamma_{\nu}^{21}(p_2,p_1)=e^{\beta\mu/2}\cos
\theta_{p_2}\sin \theta_{p_2}[S_F^{-1}(p_2)-S_F^{*-1}(p_2)]-(p_2\rightarrow p_1).
\end{equation}%%(24)%%
To tree-diagram order, we will have
$$\Gamma_{\nu}^{11}(p_2,p_1)=-\Gamma_{\nu}^{22}(p_2,p_1)=\gamma_{\nu},$$ $$
  S_F^{-1}(p)=S_F^{*-1}(p)=\not\!{p}-m, $$
and the thermal WTI (21) becomes
\begin{equation}
\left(\matrix{q^{\nu}\gamma_{\nu}&0\cr
              0 & -q^{\nu}\gamma_{\nu}\cr}\right)=
\left(\matrix{\not\!{p_2}- \not\!{p_1}&0\cr
              0& -(\not\!{p_2}-\not\!{p_1})\cr}\right)
\end{equation}%%(25)%%
which is obviously valid.  In the limit of $T=\mu=0$, we only need to consider the
(11) component of Eq.(21). Since
$$\sin \theta_p=0 , \ {\rm when} \ T=\mu=0,$$
we obtain that
\begin{equation}
q^{\nu}\Gamma_{\nu}^{11}(p_2,p_1)=S_F^{-1}(p_2)-S_F^{-1}(p_1)
\end{equation}%%(26)%%
which, with $\Gamma_{\nu}^{11}(p_2,p_1)$ being identified with the physical vertex,
exactly reproduces the standard form of the WTI at $T=\mu=0$. \\
\indent In the general case of finite $T$ and $\mu$, we see that Eq.(22) contains
the thermal correction terms, i.e. the last two terms in its right-handed side, and
Eq.(24) represents definite thermal constraints on $\Gamma_{\nu}^{12}(p_2,p_1)$ and
$\Gamma_{\nu}^{21}(p_2,p_1)$. These corrections and constraints exist only at finite
$T$ and $\mu$ and always related to a factor  $[S_F^{-1}(p)-S_F^{*-1}(p)]$, i.e. the
imaginary part of $S_F^{-1}(p)$.  Since $iS_F(p)$ is the complete Feynman propagator
for fermion, so $S_F^{-1}(p)$ will contain complete fermion self-energy and could
have an imaginary part. If we write
\begin{equation}
iS_F(p)=\frac{i}{A(p^2)\not\!{p}-B(p^2)+i\varepsilon},
\end{equation}%%(27)%%
then the factor
\begin{equation}
S_F^{-1}(p)-S_F^{*-1}(p)=[A(p^2)-A^*(p^2)]\not\!{p}-[B(p^2)-B^*(p^2)]
\end{equation}%%(28)%%
will be related to only the imaginary parts of the functions $A(p^2)$ and $B(p^2)$.
Therefore, Eqs.(22)-(24) will be thermal constraints on the imaginary part of
$S_F^{-1}(p)$. When one researches into the physical processes involving the
imaginary part of $S_F^{-1}(p)$ or say, fermion self-energy, e.g. discussing the
decay rate of particle and Landau damping etc. \cite{kn:4,kn:13}, such constraints
would certainly be quite important and must be considered carefully, even the
calculations are conducted only up to a given order of perturbation theory.  On the
other hand, for some physical processes which do not relate to the imaginary part of
$S_F^{-1}(p)$ e.g. dynamical chiral symmetry breaking and restoration, one only
needs to deal with a real mass function of fermion which can come from
$B(p^2)/A(p^2)$. In this case, we can assume that $S_F^{-1}(p)$ is a real function,
i.e.
\begin{equation}
S_F^{-1}(p)-S_F^{*-1}(p)=0, \ {\rm for \ any} \ p
\end{equation}%%(29)%%
or by Eq.(28), equivalently
\begin{equation}
A(p^2)=A^*(p^2), B(p^2)=B^*(p^2).
\end{equation}%%(30)%%
As a result we will have
$$\Gamma_{\nu}^{11}(p_2,p_1)=-[\Gamma_{\nu}^{22}(p_2,p_1)]^*
=-\Gamma_{\nu}^{22}(p_2,p_1)\equiv\Gamma_{\nu}(p_2,p_1)
$$
and the thermal WTI (21) becomes
\begin{equation}
\left(\matrix{q^{\nu}\Gamma_{\nu}(p_2,p_1)&0 \cr
 0&-q^{\nu}\Gamma_{\nu}(p_2,p_1)\cr}\right)=
\left(\matrix{S_F^{-1}(p_2)-S_F^{-1}(p_1),& 0\cr
 0& -[S_F^{-1}(p_2)-S_F^{-1}(p_1)]\cr}\right),
\end{equation}%%(31)%%
which is essentially identical to the WTI (26) at $T=\mu=0$. However, we emphasize
that only if $S_F^{-1}(p)$ is a real function, the WTI at finite $T$ and $\mu$ is
just reduced to its form at $T=\mu=0$. \\
\indent The above discussions indicate that, when one researches the Schwinger-Dyson
equation of fermion self-energy and the effective potential for complete fermion
propagator at finite $T$ and $\mu$ in the real-time formalism of thermal field
theory, one will be able to use the thermal WTI in its form at $T=\mu=0$. For
example, for the 1-type vertex of physical fields, the corresponding thermal WTI can
be written from Eq.(31) and expressed by
\begin{equation}
(p_2-p_1)^{\nu}\Gamma_{\nu}(p_2,p_1)=A(p_2^2)\not\!{p}_2-B(p_2^2)-A(p_1^2)\not\!{p}_1+B(p_1^2).
\end{equation}%%(32)%%
To satisfy Eq.(32), a possible $\Gamma_{\nu}(p_2,p_1)$ is
\begin{equation}
\Gamma_{\nu}(p_2,p_1)=\gamma_{\nu}+\frac{(p_2-p_1)_{\nu}}{(p_2-p_1)^2}
\{[A(p_2^2)-1]\not\!{p}_2-[A(p_1^2)-1]\not\!{p}_1-B(p_2^2)+B(p_1^2) \},
\end{equation}%%(33)%%
where an unspecified transverse pieces have assumedly been neglected. Such selection
given by Eq.(33) could make practical calculations be simplified greatly if the
Landau gauge is taken. In fact, in this gauge,  the factors in thermal gauge field
propagator which are coupled to $\Gamma_{\nu}(p_2,p_1)$ will have the forms
\begin{equation}
D^{\nu\mu}(p_2-p_1)\sim \Delta^{\nu\mu}(p_2-p_1)\frac{-i}{(p_2-p_1)^2+i\varepsilon},
\;\Delta^{\nu\mu}(p_2-p_1)=g^{\nu\mu}-\frac{(p_2-p_1)^{\nu}(p_2-p_1)^{\mu}}
{(p_2-p_1)^2+i\varepsilon}
\end{equation}%%(34)%%
and corresponding complex conjugate $D^{\nu\mu*}(p_2-p_1)$ and
$\Delta^{\nu\mu*}(p_2-p_1)$. Thus we obtain that
\begin{equation}
\Gamma_{\nu}(p_2,p_1)\left\{\matrix{\Delta^{\nu\mu}(p_2-p_1)\cr
                                    \Delta^{\nu\mu*}(p_2-p_1)\cr}\right.
                                    =\gamma_{\nu}\left\{\matrix{\Delta^{\nu\mu}(p_2-p_1)\cr
                                    \Delta^{\nu\mu*}(p_2-p_1)\cr}\right.,
\end{equation}%%(35)%%
i.e. effectively only a tree vertex $\gamma_{\nu}$ is left. Physically this is a
well-known result: in the Landau gauge, the gauge field propagator is transverse and
the effective coupling will be independent of the longitudinal sector of
$\Gamma_{\nu}(p_2,p_1)$. \\
\indent In conclusion, it has been shown that, based on vectorial current
conservation and canonical operator approach of fields, we can give a simple proof
of the Ward-Takahashi identity at finite temperature and finite chemical potential
for complete vectorial vertex and complete fermion propagator which is rigorous in
QED and approximate in QCD where the ghost effect in fermion sector has been
neglected. The results are expressed in thermal matrix form in the real-time thermal
field theory. All the thermal corrections to the WTI only relate to the imaginary
part of the inverse complete Feynman propagator for fermion and will play important
role in the physical processes associated with the imaginary part of fermion
self-energy. When the inverse fermion's Feynman propagator is assumed to be real,
the thermal WTI will essentially be reduced to its form at zero temperature and zero
chemical potential thus one can use it in the latter's form.  At this point, a practical
example is also mentioned. \\

\end{document}